\documentclass[aps,print,showpacs,twocolumn]{revtex4-1}
\usepackage{amsfonts}
\usepackage{amsmath}
\usepackage{amssymb}
\usepackage{graphicx}

\DeclareMathOperator{\sech}{sech}
\begin{document}

\title{Matter rogue wave in Bose-Einstein condensates with attractive atomic
interaction}
\author{Lin Wen$^{1}$, Lu Li$^{2}$, Zai-Dong Li$^{3}$, Shu-Wei Song$^{1}$,
Xiao-Fei Zhang$^{1}$, and W.M. Liu$^{1}$}

\affiliation{$^{1}$Beijing National Laboratory for Condensed Matter Physics, Institute of
Physics, Chinese Academy of Sciences, Beijing 100080, China.}
\affiliation{$^{2}$Institute of Theoretical Physics, Shanxi University,
Taiyuan, 030006, China}
\affiliation{$^{3}$Department of Applied Physics and School of Information Engineering,
Hebei University of Technology, Tianjin 300401, China}

\begin{abstract}
We investigate the matter rogue wave in Bose-Einstein Condensates with
attractive interatomic interaction analytically and numerically. Our results
show that the formation of rogue wave is mainly due to the accumulation of
energy and atoms toward to its central part; Rogue wave is unstable and the
decay rate of the atomic number can be effectively controlled by modulating
the trapping frequency of external potential. The numerical simulation demonstrate that even a small periodic perturbation with small modulation frequency can induce the generation of a
near-ideal matter rogue wave. We also give an experimental protocol to
observe this phenomenon in Bose-Einstein Condensates.
\end{abstract}


\maketitle

\section{Introduction}

The dynamics of Bose-Einstein condensates (BECs) at ultralow temperature are described
well by Gross-Pitaevskii (GP) equation \cite{GP}, in
which the nonlinearity is arose from interatomic interactions characterized
by the s-wave scattering length. Recently experiments have
demonstrated that ¡°tuning¡± of the effective scattering length, including a
possibility to change its sign, can be achieved by using the
so-called Feshbach resonance technique \cite{Fechbach_B}. In particular, the experimental realization of BECs
in dilute quantum gases have opened the floodgate in the field of atom optics and condensed matter physics \cite{BECs}. At the same time,
the collective excitation of matter waves in BECs has also drawn a great deal of interest to explore the dynamics of BECs deeply from both
experimental and theoretical perspectives, such as matter wave solitons \cite{dark1,Soliton,breather,Soliton1,Soliton2,Soliton3}, periodic waves \cite{periodic wave}, shock waves \cite{shock wave},
vortex \cite{votex} and necklaces \cite{necklace}.
However, to our knowledge less attention have been paid to the matter rogue wave which is a fundamental and novel nonlinear excitation in BECs.

Rogue waves from ocean are that their heights, from crest to trough,
are more than about twice the significant wave height  \cite{rogue}. They appear without
any warning and disappear without the slightest trace. Owing to severe
environment and high risk in ocean, the systematic study of rogue waves
become so difficult that the necessary conditions and physical mechanism of
their generation are not sufficiently well understood. Recently, theoretical
studies have shown that the rogue wave phenomenon can be explained well by
nonlinear theories \cite{Gener_Rogue,Peregrine}, and the various possible
formative mechanisms have been discussed, such as the modulation instability
in one dimension \cite{1,2}, nonlinear spectral instability \cite{3} and in
two-dimensional crossings \cite{5}. Furthermore, the rogue waves phenomenon have been observed experimentally in variety of physical systems including optical fibers \cite{optics rogue,NatPhys}, arrays
of optical waveguide \cite{optical rogue} and capillary waves \cite{capillary waves}.

As a nonlinear physical system with similar nonlinear characteristics, BECs
can support the interesting rogue waves and allow us to understand deeply the nature and the
dynamics of rogue waves in laboratory conditions. The management of Feshbach
resonance for nonlinearity and a tunable atomic trapping potential also
provide us with a powerful tool for manipulating rogue wave. This enable
BECs to have more advantages for investigating rogue wave than other
physical media. In this paper, we mainly investigate the matter rogue wave
of Peregrine type with the emphasis on its formative mechanism in BECs. We firstly obtain the exact rogue
wave solution of the GP equation with time-dependent attractive atomic interaction in an expulsive parabolic potential. By analyzing the
atomic number density distribution in the rogue wave
against the background, the formative mechanism of matter rogue
wave can be clarified that the accumulation of energy and
atoms toward to its central part. Rogue wave can not
keeps dynamic stability and the decay rate of the atomic
number can be effectively controlled by modulating the
trapping frequency of external potential. Moreover, we
use the breather evolution in a regime approaching the
excitation of matter rogue wave as approximation to simulate
the creation of matter rogue wave numerically, which indicates
that a small periodic perturbation with small modulation
frequency can excite a near-ideal matter rogue wave. Finally, we
also give a practical and effective experimental protocol to observe this
interesting phenomenon in future BECs experiments.

\section{MATTER ROGUE WAVE SOLUTION}

Under the mean-field level, the evolution of macroscopic wave function of
BECs obey the 3D GP equation \cite{BECs}. For a cigar-shaped condensate at a
relatively low density, when the energy of two body interactions is much
less than the kinetic energy in the transverse direction, the system can
become effectively quasi-one-dimensional regime with time-dependent
attractive interaction in an expulsive potential \cite{1D},
\begin{figure}[tbp]
\includegraphics[width=7.5cm]{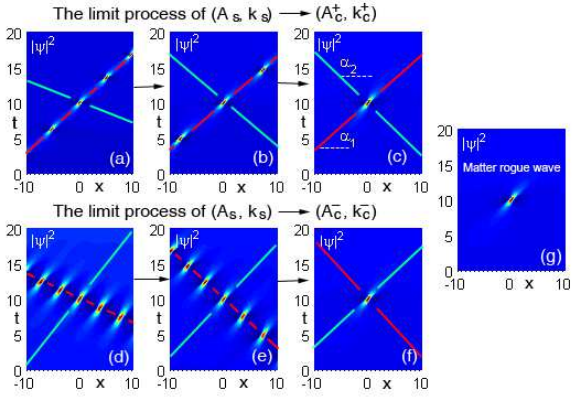}
\caption{(Color online) The asymptotic processes from Eq. (\protect\ref{sol}%
) to Eq. (\protect\ref{Pere sol}) in the limitation of $\left(
A_{s},k_{s}\right) \rightarrow \left( 2A_{c},k_{c}\right) $. As the bright
soliton amplitude $A_{s}$ and wave number $k_{s}$ approaching $(2A_{c},k_{c})
$, the spatio-temporal separation between adjacent peaks gradually increases
shown in Fig. 1(a-f), where the parameters are (a) $A_{s}=2.4,k_{s}=1.2.$
(b) $A_{s}=2.2,k_{s}=1.2.$ (c) $A_{s}=2.01,k_{s}=1.01.$ (d) $%
A_{s}=1.6,k_{s}=0.8.$ (e) $A_{s}=1.8,k_{s}=0.8.$ (f) $A_{s}=1.99,k_{s}=0.99$%
. Fig. 1(g) represents the exact matter rogue wave solution Eq. (\protect\ref%
{Pere sol}) with $A_{s}=2,k_{s}=1$. Other parameters are $\protect\lambda %
=0.01,t_{0}=10,k_{c}=A_{c}=1.$ The red and green lines represent the sloping
lines $V_{\protect\theta \text{ }}$and $V_{\protect\alpha }$, respectively.
}
\end{figure}

\begin{equation}
i\frac{\partial \psi }{\partial t}+\frac{1}{2}\frac{\partial ^{2}\psi }{%
\partial x^{2}}+a(t)\left\vert \psi \right\vert ^{2}\psi +\frac{1}{2}\lambda
^{2}x^{2}\psi =0,  \label{model}
\end{equation}%
where the aspect ratio reads $\lambda =\left\vert \omega _{0}\right\vert
/\omega _{\bot }\ll 1$, coordinate $x$ and time $t$ are measured in units $%
a_{\bot }$ and $1/\omega _{\bot }$ with $a_{\bot }=\sqrt{\hbar /m\omega
_{\bot }}$ ($m$ is the atomic mass) and $a_{0}=\sqrt{\hbar /m\omega _{0}}$
the linear oscillator lengths in the transverse and cigar-axis directions,
respectively. $\omega _{\bot }$ and $\omega _{0}$ are corresponding harmonic
oscillator frequencies. The nonlinear coefficient $a(t)$ is defined as $%
a(t)=|a_{s}(t)|/a_{B},$ where $a_{s}(t)$ is so-called $s$-$wave$ scattering
length. Corresponding to the real BECs experiment \cite%
{Soliton3}, the bright soliton can be created for $^{7}Li$ by tuning the
scattering length continuously in an expulsive potential with $\omega _{\bot
}$ $=$ $2\pi \times 700Hz$ and $\omega _{0}=2i\pi \times 7Hz$. So we can
choose the nonlinear coefficient in the form of $a\left( t\right) =\exp
[\lambda (t-t_{0})]$ manipulated by Feshbach resonance technique \cite{time-dependent nonlinearity}, where $t_{0}$ represents an arbitrary real constant
determining the initial scattering length $\left\vert a_{s}\left( t=0\right)
\right\vert =a_{B}e^{-\lambda t_{0}}$.

To obtain the exact solution of Eq. (\ref{model}), we introduce the
transformation $\psi=q(X,T)\exp[\lambda(t-t_{0})/2-i\lambda x^{2}/2]$ with
the coordinate transformations $X=e^{\lambda(t-t_{0})}x$ and $T=[e^{2\lambda
(t-t_{0})}-1]/\left( 2\lambda \right) $. Then Eq. (\ref{model}) can reduce
to the standard nonlinear Schr\"{o}dinger equation, and the solution of Eq. (%
\ref{model}) constructed on continuous wave (cw) background $%
\psi_{cw}=A_{c}e^{i\varphi}$ can be obtained as follows \cite{Lilu},

\begin{align}
\psi & =\left( A_{c}+A_{s}\frac{\chi \cosh \theta+\cos \alpha}{\cosh \theta
+\chi \cos \alpha}+iA_{s}\frac{\eta \sinh \theta+\delta \sin \alpha}{\cosh
\theta+\chi \cos \alpha}\right)  \label{sol} \\
& \times \exp(i\varphi),  \notag
\end{align}
with
\begin{equation*}
\chi=\frac{-2A_{c}A_{s}}{A_{s}^{2}+M_{R}^{2}},\text{ }\eta=\frac{-2A_{c}M_{R}%
}{A_{s}^{2}+M_{R}^{2}},\text{ }\delta=\frac{M_{I}}{A_{s}}\allowbreak,
\end{equation*}
where $\theta=M_{I}X-[A_{s}M_{R}+\left( k_{c}+k_{s}\right) M_{I}]T/2$, $%
\alpha=M_{R}X-[(k_{c}+k_{s})M_{R}-A_{s}M_{I}]T/2,$ $\varphi=\varphi
_{c}-\lambda x^{2}/2-i\lambda(t-t_{0})/2,$ $%
\varphi_{c}=k_{c}X+(A_{c}^{2}-k_{c}^{2}/2)T$ and $M_{R}+iM_{I}=\sqrt{%
(k_{c}-k_{s}-iA_{s})^{2}+4A_{c}^{2}}$. The subscripts $R$ and $I$ denote the
real and imaginary part of $M$, respectively.

In general, some excited state solutions, such as cw
wave and bright soliton on the background of ground state, can be recovered
successfully from Eq. (\ref{sol}). Firstly, when the cw background amplitude
vanishes ($A_{c}=0$), Eq. (\ref{sol}) reduces to the bright
soliton solution $\psi _{sol}=A_{s}e^{i\varphi _{s}}\sech \theta _{s}$ with
varying amplitude $A_{s}e^{\lambda (t-t_{0})}$ and group volecity $%
v_{s}=k_{s}\cosh [\lambda (t-t_{0})]$ \cite{Serkin}, where $\theta
_{s}=A_{s}(X-k_{s}T)$ and $\varphi _{s}=k_{s}X+(A_{s}^{2}-k_{s}^{2})T/2$,
and $k_{s}$ represents the wave number of bright soliton.
Secondly, when the initial amplitude of bright soliton vanishes ($A_{s}=0$),
Eq. (\ref{sol}) reduces to cw background solution $\psi
_{cw}=A_{c}e^{i\varphi }$ with varying group velocity $v_{c}=k_{c}\cosh
[\lambda (t-t_{0})]$, where $k_{c}$ is the wave number of
cw background. So the exact solution Eq. (\ref{sol}) represents a bright soliton
embedded in a cw background\ field. It should be pointed that the parameters $A_{c},A_{s},k_{c}$ and $k_{s}$ are
derived from the mathematical construction of Eq. (2), the numerical
values of these parameters can be chosen freely, so we assume that these
parameters are real constants without loss of generality. Furthermore, the dynamical evolution of solution Eq. (\ref{sol}) are
shown in Fig. 1(a-f). From Fig. 1 we observe that the solution in Eq. (\ref%
{sol}) commonly exhibits a breather characteristic and a time periodic
modulation of the soliton amplitude, which can be regarded as the results of
the interaction between the localized process of cw\textit{\ }background
along the slope direction $V_{\alpha }$ and the periodization process of
bright soliton along the slope direction $V_{\theta }$, i.e., the bright soliton undergo periodic energy and atoms exchange with cw background, where $V_{\theta }$
and $V_{\alpha }$ represent the lines $M_{I}X-[A_{s}M_{R}+\left(
k_{c}+k_{s}\right) M_{I}]T/2=0$ and $%
M_{R}X-[(k_{c}+k_{s})M_{R}-A_{s}M_{I}]T/2=0$ on the space-time plane as
shown in Fig. 1, respectively.
\begin{figure}[tbp]
\includegraphics[width=8cm]{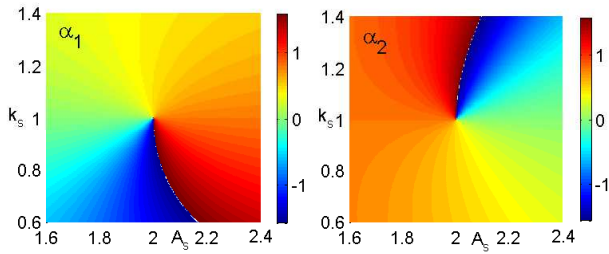}
\caption{(Color online) The plots for the evolutions of $\protect\alpha _{1}$
and $\protect\alpha _{2}$ as the function of $A_{s}$ and $k_{s}$ during the
limit processes, where $k_{c}=1$ and $A_{c}=1$. The two angles $\protect%
\alpha _{1}$ and $\protect\alpha _{2}$ as the function of $A_{s}$ and $k_{s}$
are not continuous at the limit point.}
\end{figure}

Through above analysis we can see that when $k_{c}=k_{s}$, the bright soliton must move with the same group velocity as
that of cw background which implies that the bright soliton and cw
background are in \textquotedblleft resonant state" in space. The critical point $A_{s}^{2}=4A_{c}^{2}$
forms the dividing line between modulation instability process ($%
A_{c}^{2}>A_{s}^{2}/4$) and the periodization process of bright soliton ($%
A_{c}^{2}<A_{s}^{2}/4$) under the resonant condition $k_{c}=k_{s}$ \cite{Lilu}, which
means that the relative initial intensity between the cw background and
bright soliton determine the different physical behaviors of the solution
Eq. (\ref{sol}). Especially, with the limit conditions of $k_{c}=k_{s}$ and $%
A_{s}^{2}=4A_{c}^{2}$, we have
\begin{equation}
\psi _{R}=\left[ \frac{4+i8A_{c}^{2}T}{1+4A_{c}^{4}T^{2}+4A_{c}^{2}\left(
X-k_{c}T\right) ^{2}}-1\right] A_{c}e^{i\varphi },  \label{Pere sol}
\end{equation}%
which represents a localized matter wave with the maximal amplitude $%
A_{P}=3A_{c}$ in BECs, and its dynamical evolution is shown in Fig. 1(g).
Interestingly, the exact solution Eq. (\ref{Pere sol}) displays the typical
rogue wave characteristics of Peregrine type that a localized breather characteristic with
only a single hump both in space and time, which indicates that the
localized wave is captured completely at $x=0$ and $t=t_{0}$ by the cw
background  \cite{Peregrine}. So far, such solution Eq. (\ref{Pere sol}) has been conjectured
to be a prototype of oceanic rogue waves.

\section{Dynamics of MATTER ROGUE WAVE}
In order to better clarify the formative mechanism of rogue wave solution in
Eq. (\ref{Pere sol}), we firstly investigate the asymptotic processes of Eq. (\ref%
{sol}) to Eq. (\ref{Pere sol}) in the limit processes $%
(A_{s},k_{s})\rightarrow (2A_{c},k_{c})$ by fixing the numerical values of cw background
amplitude $A_{c}$ and wave number $k_{c}$. From Fig. 1(a-f), we can observe clearly that the
spatio-temporal separation between adjacent peaks gradually increases  as the bright soliton amplitude $%
A_{s}$ and wave number $k_{s}$ approaching $(2A_{c},k_{c})$, which leads to a greater spatio-temporal localization in Eq. (\ref%
{sol}). Furthermore, the parameters $\tan \alpha _{1}=2/$($%
k_{c}+k_{s}+A_{s}M_{R}/M_{I}$) and $\tan \alpha
_{2}=2/(k_{c}+k_{s}-A_{s}M_{I}/M_{R})$ shown in Fig. 1(c) represent the
slope of the lines $V_{\theta }$ and $V_{\alpha }$ at $x=0$ and $t=t_{0}$,
respectively. When the values of $A_{s}$ and $k_{s}$ approaches to the
critical point, $V_{\theta }$ and $V_{\alpha }$ gradually turn to a relative
fixed direction associated with the maximal spatio-temporal localization in
Eq. (\ref{sol}). However, the two angles $\alpha _{1}$ and $\alpha _{2}$ as
the function of $A_{s}$ and $k_{s}$ are not continuous at the limit point
shown in Fig. 2, i.e., $\alpha _{1}$ and $\alpha _{2}$ do not exist
limitation, which play the important role to describe that the solution Eq. (%
\ref{sol}) is localized along $V_{\theta }$ and $V_{\alpha }$ at the limit
point. Especially, rogue wave solution in Eq. (\ref{Pere sol}) can be
considered as a transition state between the modulation instability process (%
$A_{s}\rightarrow 2A_{c}^{-}$) and the periodization process of the bright
soliton ($A_{s}\rightarrow 2A_{c}^{+}$) under the resonant condition $k_{c}=k_{s}$.

As shown in the following, the formation of rogue wave can be clarified by
the atomic number density distribution against the background defined as $%
\rho \left( x,t\right) =\left\vert \psi _{R}\left( x,t\right) \right\vert
^{2}-\left\vert \psi _{R}\left( x=\pm \infty ,t\right) \right\vert ^{2}$.
With Eq. (\ref{Pere sol}), we have
\begin{equation}
\rho \left( x,t\right) =\frac{%
8A_{c}^{2}+32A_{c}^{6}T^{2}-32A_{c}^{4}(X-k_{c}T)^{2}}{\left[
1+4A_{c}^{4}T^{2}+4A_{c}^{2}(X-k_{c}T)^{2}\right] ^{2}}e^{\lambda \left(
t-t_{0}\right) },  \label{exch}
\end{equation}%
and the time-independent integral $\int_{-\infty }^{+\infty }\rho (x,t)dx=0$%
. From the condition $\rho (\pm 1/\left( 2A_{c}\right) ,t_{0})=0$ and $\rho
(0,t_{0})=8A_{c}^{2}$, one can define the spatial width of the hump part in
rogue wave as $1/A_{c}$. At the fixed time $t=t_{0}$, we have integral $%
\int_{-1/(2A_{c})}^{1/(2A_{c})}\rho (x,t_{0})dx=4A_{c}$ and $\int_{-\infty
}^{-1/(2A_{c})}\rho (x,t_{0})dx+\int_{1/(2A_{c})}^{\infty }\rho
(x,t_{0})dx=-4A_{c}$. These results demonstrates clearly that for the
attractive interatomic interaction, the generation of rogue wave with
stronger breather characteristic is mainly due to the accumulation of energy
and atoms toward to its central part. The time-independent area relation
shown in Fig. 3(a), i.e., $S_{1}+S_{2}=S_{3}$, means that the loss of atoms
in background completely transfer to the hump part of rogue wave. The
forthcoming fundamental problem is that how rogue wave gather atoms and
energy toward to its central part from the background. To this purpose, we
investigate the exact number of atomic exchange between rogue wave and
background which has the form
\begin{figure}[tbp]
\includegraphics[width=8cm]{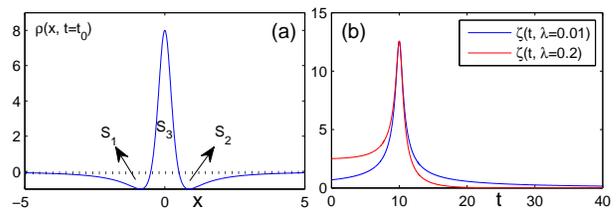}
\caption{(Color online) (a) The atomic number density distribution in matter
rogue wave at fixed time point. (b) The atomic exchange between matter rogue
wave and background. The parameters are $A_{c}=k_{c}=1$ and $t_{0}=10.$}
\end{figure}
\begin{equation}
\zeta \left( t\right) =\int_{-\infty }^{\infty }\left\vert \psi _{R}\left(
x,t\right) -\psi _{R}\left( \pm \infty ,t\right) \right\vert ^{2}dx=\frac{%
4\pi A_{c}}{\sqrt{1+4A_{c}^{4}T^{2}}}.  \label{densi}
\end{equation}%
From the above expression, we can see that $\zeta (t)$ is time-aperiodic
which is different from periodic exchange of atoms between the bright
soliton and the cw background in Eq. (\ref{sol}). As shown in Fig. 3(b), the
atoms in background is gathered to the central part when $t<t_{0}$, which
leads to the generation of a hump with two fillisters on the background along
the space direction. The maximal peak of the hump and the deepest fillisters
occur at $t=t_{0}$. However, the atoms in the hump start to spread into the
fillisters when $t>t_{0}$. Therefore, the hump gradually
decay which verifies that the rogue wave is only one oscillation in time and
displays a unstable dynamical behavior.

Considering the dynamics of rogue wave in the background, on the one hand, a necessary
condition for realizing rogue wave in experiment is that the scale of rogue
wave must be very small compared with the length of the background of BECs.
In the real experiments \cite{Soliton2}, the length of the background of
BECs is at least $370\mu m$. At the same time, in Fig. 3(a), the actual
width of rogue wave is about $10a_{\bot}=14\mu m\ll370\mu m$ $($a unity of
coordinate corresponds to $a_{\bot}=\sqrt{\hbar/m\omega_{\bot}}$ $=1.4\mu m)$%
. Thus the rogue wave is observable experimentally. On the other hand, the decay
rate of atoms in rogue wave can be controlled effectively by modulating the
trapping frequency of external potential shown in Fig. 3(b). The decay time $%
t_{d}$ is about $8.05ms$ for $\lambda=0.01$ ($\omega_{\perp}=2\pi
\times700Hz $ and $\omega_{0}=2i\pi \times7Hz$ originate from the experiment
\cite{Soliton3}), while $t_{d}\approx$ $2.30ms$ for $\lambda=0.2$ ($%
\omega_{\perp}=2\pi \times700Hz$ and $\omega_{0}=4i\pi \times70Hz$), which
demonstrate that a trap with small trapping frequency is conducive to the
observation of matter rogue wave in BECs experiments.
\begin{figure}[ptb]
\includegraphics[width=8cm]{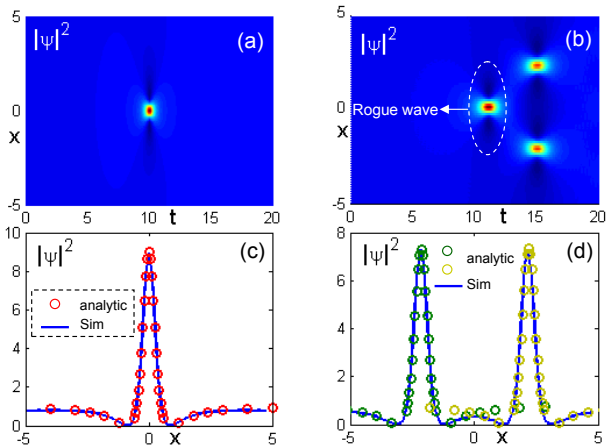}
\caption{(Color online) (a) Evolution of the exact rogue wave solution Eq. (%
\protect\ref{Pere sol}), where $A_{c}=1,k_{c}=0,\protect\lambda=0.01$ and $%
t_{0}=10$. (b) Evolution of the numerical solution of rogue wave with the
initial condition Eq. (\protect\ref{MI}), where $A_{s}=1.92$, $\protect%
\lambda=0.01,t_{0}=10.$ (c) and (d), The comparison of intensity profile
between near-ideal rogue wave (in (c) with solid blue line) or sub-rogue
wave pair (in (d) with solid blue line) and ideal rogue wave (circles).}
\end{figure}

A question arises about the possibility of the creation of such a matter
rogue wave experimentally. Generally, the excitation of rogue wave can be
recovered by means of the numerical simulation with some particular initial
conditions \cite{Initial_con}. From the experimental points of view, optimal
initial conditions are not only conducive to the experimental preparation,
but it is also useful for understanding the necessary conditions and
physical mechanism of the generation of rogue wave. In what follows, we will
look for the optimal initial conditions which can excite the resemble
physical behavior of rogue wave.

By comparing the Fig. 1(c) and (f) with Fig. 1(g), Eq. (\ref{sol}) with
infinity oscillation period is a very good approximation of rogue wave
solution Eq. (\ref{Pere sol}), this parameter regime yields characteristic
rogue wave features in the spatio-temporal envelope even though the ideal
rogue wave solution exists only asymptotically in the limitation of $%
A_{c}=2A_{s}$ and $k_{c}=k_{s}$. Based on modulation instability mechanism in ultracold
atom system \cite{MI}, we consider the
case of $A_{c}^{2}>A_{s}^{2}/4$ with $k_{s}=k_{c}=0$ corresponding to the
modulation instability process of cw background. In this case, we can take $%
T\approx-t_{0}$ at $t=0$ for a very small quantity $\lambda$, and the
suitable values of $A_{s}$ and $A_{c}$ can ensure $\theta \approx
t_{0}A_{s}M_{R}/2$ to be so large that $\kappa=e^{-\theta}$ is a small
quantity. By linearizing the initial value with the small quantity $\kappa$
in Eq. (\ref{sol}) we get
\begin{equation}
\psi \left( x,0\right) =\left( \sigma+\gamma \kappa \cos \alpha_{0}\right)
e^{i\varphi \left( x,0\right) },  \label{MI}
\end{equation}
where $\sigma=(2A_{c}^{2}-A_{s}^{2}-iA_{s}M_{R})/(2A_{c}),\gamma=A_{s}M_{R}%
\left( M_{R}-iA_{s}\right) /(2A_{c}^{2})$ and $\alpha_{0}=M_{R}e^{-\lambda
t_{0}}x,$ $\varphi_{c}\approx-A_{c}^{2}t_{0}$ and the modulation frequency
of initial condition Eq. (\ref{MI}) is $\Omega =M_{R}e^{-\lambda t_{0}}$.
The solution of initial value problem associated with Eq. (\ref{model}) with
initial condition Eq. (\ref{MI}) can be described well by the exact solution
Eq. (\ref{sol}) \cite{Lilu}. For the creation of rogue wave, we require the
value of $A_{s}/2$ to approach $A_{c}$ associated with a very small
modulation frequency in Eq. (\ref{MI}). The numerical
results are shown in Fig. 4(b), which demonstrates that a small periodic
perturbation with a very small modulation frequency can induce a near-ideal
rogue wave localization, whose profile is basically consistent with the
ideal theoretical limit solution Eq. (\ref{Pere sol}) as shown in Fig. 4(a)
and (c). However, the obvious difference is that owing to the actions of
modulation instability and instability of rogue wave, the initial near-ideal
rogue wave can break up into two lower amplitude but equally strongly
localized sub-rogue wave, and each sub-rogue wave itself exhibits ideal
rogue wave characteristics as shown in Fig. 4(d), which agrees well with the
optical experimental conclusions \cite{optics rogue,NatPhys}. As a result, a
small initial periodic perturbation with a small modulation frequency can
induce the generation of a near-ideal rogue wave by the
modulation instability mechanism in BECs.

Inspired by the experiments \cite{Soliton1,Soliton2,Soliton3}, we can design
the following experimental steps to observe the interesting rogue wave
phenomenon in BECs: (i) Creating a condensates of $^{7}Li$ with total number
of particles $N\approx \times 10^{3}$ and continuous wave phase
distribution by using quantum phase imprinting technique and amplitude
engineering; (ii) Loading the condensates into a slightly expulsive harmonic
potential with the parameters $\omega _{\perp }=2\pi \times 700Hz$ and $%
\omega _{0}=2i\pi \times 7Hz$, and ramping up the scattering length in the
form of $a_{s}\left( t\right) =-0.9a_{B}e^{\lambda t};$ (iii) Optimal
initial state Eq. (\ref{MI}) can be produced by imprinting a periodic
perturbation laser with very small modulating frequency onto
condensates. The main effect of this expulsive term is that the center of
the condensates accelerates along the longitudinal direction. In addition, a
crucial question is that since we require the scattering length to change
over time in above experimental protocol, we must ensure the validity of
quasi-one-dimensional regime and avoid the collapse of condensates with
attractive interaction. In other words, we must ensure that the energy of
two body interactions is much less than the kinetic energy in the transverse
direction, i.e., $\varepsilon ^{2}\sim N\left\vert a_{s}\right\vert
/a_{0}\ll 1$. With initial scattering length $a_{s}(t=0)=-0.9a_{B}$, we can
obtain $\varepsilon ^{2}\approx 0.03\ll 1$. After about 50 dimensionless
units of time, the scattering length becomes $|a_{s}(t)|=1.4a_{B}$
corresponds to $\varepsilon ^{2}\approx 0.057$ $\ll 1$. So the validity of
quasi-one-dimensional system can be maintained well. Finally, we emphasize
that the interesting phenomenon of rogue wave can be observed within current
experimental capability.

\section{Conclusions}

In conclusion, we have investigated the formative mechanism of matter rogue
wave in BECs with time-dependent interaction in an expulsive parabolic
potential, analytically and numerically. The generation of rogue wave with
stronger breathing characteristic is mainly due to the accumulation of
energy and atoms toward to its central part. Rogue wave can not keeps
dynamic stability because of the aperiodic exchange of energy and atoms with
the background; The decay rate of the number of atoms in rogue wave can be
controlled effectively by modulating the trapping frequency of the external
potential. Our numerical results show that a small periodic perturbation
with a smaller modulating frequency can induce the generation of the
near-ideal rogue wave. Finally, we emphasize that the interesting phenomenon
of rogue wave in BECs can be observed experimentally. Our work may facilitate the
deeper studies of hydrodynamic rogue wave.

The authors thank Prof. Z. Y. Yan and B. Xiong for helpful discussions. This
work was supported by the NSFC under grants Nos. 10874235, 10934010,
61078079, 60978019, 10874038 and the NKBRSFC under grants Nos. 2009CB930701,
2010CB922904 and 2011CB921502, NSFC-RGC under grants Nos1386-N-HKU748/10,
and the Hundred Innovation Talents Supporting Project of Hebei Province of
China under Grant No. CPRC014.

\bigskip

\bigskip

\bigskip

\bigskip

\bigskip

\bigskip

\bigskip

\bigskip

\bigskip

\bigskip

\bigskip


\begin{thebibliography}{99}
\bibitem{GP} L. Pitaevskii and S. Stringari, \textit{Bose-Einstein
Condensation} (Oxford University Press, New York, 2003).

\bibitem{Fechbach_B} J. L. Roberts \textit{et al}., Phys. Rev. Lett. \textbf{%
81}, 5109 (1998); J. Stenger \textit{et al}., Phys. Rev. Lett. \textbf{82},
2422 (1999).

\bibitem{BECs} F. Dalfovo, S. Giorgini, L. P. Pitaevskii, and S. Stringari,
Rev. Mod. Phys. \textbf{71}, 463 (1999).

\bibitem{dark1} A. D. Jackson \textit{et al}., Phys. Rev. A \textbf{58},
2417 (1998); A. E. Muryshev \textit{et al}., Phys. Rev. A \textbf{60}, R2665
(1999); P. O. Fedichev \textit{et al}., Phys. Rev. A \textbf{60}, 3220
(1999); Th. Busch \emph{et al.,} Phys. Rev. Lett. \textbf{87}, 010401 (2001).

\bibitem{Soliton} W. P. Zhang \textit{et al}., Phys. Rev. Lett. \textbf{72},
60 (1994); R. Dum \textit{et al}., Phys. Rev. Lett. \textbf{80}, 2972
(1998); X. F. Zhang \textit{et al}., Phys. Rev. A \textbf{77}, 023613 (2008).

\bibitem{breather} Z. X. Liang \textit{et al}., Phys. Rev. Lett.
\textbf{94}, 050402 (2005); B. Li \textit{et al}., Phys. Rev. A \textbf{78}, 023608 (2008);
 H. Saito \textit{et al}., Phys. Rev. Lett. \textbf{90}, 040403 (2003).

\bibitem{Soliton1} S. Burger \textit{et al}., Phys. Rev. Lett. \textbf{83},
5198 (1999); J. Denschlag \textit{et al}., Science \textbf{287}, 97 (2000).

\bibitem{Soliton2} K. E. Strecker, G. B. Partridge, A. G. Truscott, and R.
G. Hulet, Nature (London) \textbf{417}, 150 (2002).

\bibitem{Soliton3} L. Khaykovich, F. Schreck, G. Ferrari, T. Bourdel, J.
Cubizolles, L. D. Carr, Y. Castin, and C. Salomon, Science \textbf{296},
1290 (2002).

\bibitem{periodic wave} F. Kh. Abdullaev \textit{et al}., Phys. Rev. Lett. \textbf{90}, 230402 (2003).

\bibitem{shock wave} V. M. Perez-Garcia, V. V. Konotop and V. A. Brazhnyi,
Phys. Rev. Lett. \textbf{92}, 220403 (2004).

\bibitem{votex} B. P. Anderson \emph{et al.,} Phys. Rev. Lett. \textbf{86}, 2926 (2001).

\bibitem{necklace} G. Theocharis \emph{et al.,} Phys. Rev. Lett. \textbf{90}, 120403 (2003).

\bibitem{rogue} C. Kharif and E. Pelinovsky, Eur. J. Mech. B/Fluids \textbf{22}, 603
(2003); P. M\"{u}ller, Ch. Garrett, and A. Osborne, Oceanogr.
\textbf{18}, 66 (2005).

\bibitem{Gener_Rogue} A. R. Osborne, Mar. Struct. \textbf{14}, 275 (2001).

\bibitem{Peregrine} D. H. Peregrine, J. Austral. Math. Soc. \textbf{25}, 16
(1983).

\bibitem{1} K. L. Henderson, K. L. Peregrine, J. W. Dold, Wave Motion
\textbf{29}, 341 (1999).

\bibitem{2} M. Onorato, A. R. Osborne, M. Serio, S. Bertone, Phys. Rev.
Lett. \textbf{86}, 5831 (2001).

\bibitem{3} P. A. E. M. Janssen, J. Phys. Oceanogr. \textbf{33}, 863 (2003).

\bibitem{5} M. Onorato, A. R. Osborne, M. Serio, Phys. Rev. Lett. \textbf{96}%
, 014503 (2006); P. K. Shukla, I. Kourakis, B. Eliasson, M. Marklund, L.
Stenflo, Phys. Rev. Lett. \textbf{97}, 094501 (2006).

\bibitem{optics rogue} D. R. Solli \emph{et al.,} Nature \textbf{450, }1054 (2007);\ D. -I. Yeom \emph{et al.,} Nature
(London) \textbf{450}, 953 (2007).

\bibitem{NatPhys} B. Kibler \textit{et al.}, Nature Phys. \textbf{6}, 790
(2010); K. Hammani \textit{et al.}, Opt. Lett. \textbf{36, }112 (2011).

\bibitem{optical rogue} Yu. V. Bludov, V. V. Konotop, N. Akhmediev, Opt. Lett. \textbf{34}, 3015 (2009).

\bibitem{capillary waves} M. Shatz, H. Punzmann, H. Xia, Phys. Rev. Lett. \textbf{104}, 104503 (2010).

\bibitem{1D} G. Fibich \textit{et al}., Phys. Rev. Lett. \textbf{90}, 203902
(2003); P. G. Kevrekidis \emph{et al.,} Mod. Phys. Lett. B \textbf{%
18}, 173 (2004).

\bibitem{time-dependent nonlinearity} P. G. Kevrekidis \emph{et al.,} Phys. Rev. Lett. \textbf{90}, 230401 (2003).

\bibitem{Lilu} L. Li \textit{et al}., Opt. Commun.
\textbf{234}, 169 (2004); Q. Y. Li \textit{et al}., Opt. Commun. \textbf{%
283}, 3361 (2010).

\bibitem{Serkin} V. N. Serkin, A. Hasegawa, T. L. Belyaeva, Phys. Rev. Lett.
\textbf{98,} 074102 (2007).

\bibitem{Initial_con} N. Akhmediev, J. M. Soto-Crespo, and A. Ankiewicz,
Phys. Rev. A \textbf{80}, 043818 (2009).

\bibitem{MI} L. D. Carr and J. Brand, Phys. Rev. Lett. \textbf{92}, 040401 (2004); L. Salasnich, A. Parola, and L. Reatto, Phys. Rev. Lett. \textbf{91}, 080405 (2003).
\end{thebibliography}
\end{document}